\documentclass{PoS}

\usepackage{cite}

\newcommand{\bq}{\begin{eqnarray}}
\newcommand{\eq}{\end{eqnarray}}
\newcommand{\eps}{\varepsilon}

\title{NLO corrections to Z production in association with several jets}

\ShortTitle{NLO corrections to Z production in association with several jets}

\author{Daniel G\"otz\\
        PRISMA Cluster of Excellence, Johannes Gutenberg-Universit\"at Mainz\\
        E-mail: \email{goetz@uni-mainz.de}}

\author{Christian Reuschle\\
        Karlsruhe Institute of Technology, Karlsruhe\\
        E-mail: \email{christian.reuschle@kit.edu}}

\author{Christopher Schwan\\
        PRISMA Cluster of Excellence, Johannes Gutenberg-Universit\"at Mainz\\
        E-mail: \email{schwan@uni-mainz.de}}

\author{\speaker{Stefan Weinzierl}\\
        PRISMA Cluster of Excellence, Johannes Gutenberg-Universit\"at Mainz\\
        E-mail: \email{stefanw@thep.physik.uni-mainz.de}}

\abstract{
In this talk we report on first results from the NLO computation of $Z$ production in association with
five jets in hadron-hadron collisions.
The results are obtained with the help of the numerical method, where apart from the phase space
integration also the integration over the loop momentum is performed numerically.
In addition we discuss several methods and techniques 
for the improvement of the Monte Carlo integration.
         }

\FullConference{ Loops and Legs in Quantum Field Theory,\\
		 27 April - 2 May 2014 \\
		 Weimar, Germany }

\begin{document}

\section{Introduction}

The LHC experiments have reported on the measurement of $Z$ production in association with
up to seven jets \cite{Aad:2013ysa,CMS:2014gba}.
It is a challenge for the theory side to match the experimental results with
accurate and precise NLO calculations.
The NLO corrections to $Z + 0 \;\mbox{jets}$, $Z + 1 \;\mbox{jet}$, $Z + 2 \;\mbox{jets}$
have been known for a long time and have been incorporated into programs like MCFM \cite{Campbell:2002tg}.
Recently, the Blackhat collaboration reported 
on the NLO corrections to $Z + 3 \;\mbox{jets}$ and $Z + 4 \;\mbox{jets}$ \cite{Berger:2010vm,Ita:2011wn}.
In order to go towards even higher parton multiplicities a method with a good 
scaling behaviour with respect to the number of jets is required.

Methods with factorial or exponential growth are not suited for this task and one aims
for a method with polynomial growth.
For leading-order calculations it is possible to achieve a
scaling behaviour of $n^3$, where $n$ is the number of external particles of the relevant
matrix elements.
Methods with this scaling behaviour are based on recurrence relations \cite{Berends:1987me}, which recycle in a smart way already
calculated quantities and avoid in this way to calculate a quantity more than once.

In going from LO to NLO one has to face in addition loop amplitudes.
Modern methods for the virtual part achieve a polynomial growth for loop amplitudes as well.
The unitarity method \cite{Bern:1995cg,Britto:2004nc,Forde:2007mi,Ossola:2006us,Mastrolia:2008jb,Anastasiou:2006jv,Anastasiou:2006gt,Ellis:2007br,Giele:2008ve,Ellis:2008ir,Cascioli:2011va,Actis:2012qn,Badger:2012pg} scales asymptotically as $n^8$ or $n^9$, 
depending on the cache system used \cite{Ellis:2011cr,Badger:2010nx}. 
For intermediate values of $n$ a scaling behaviour of $n^6$ is observed empirically.
Within the numerical method \cite{Nagy:2003qn,Gong:2008ww,Assadsolimani:2009cz,Assadsolimani:2010ka,Becker:2010ng,Becker:2011vg,Becker:2012aq,Becker:2012nk,Becker:2012bi} discussed here one
can achieve a scaling behaviour of $n^3$.

\section{The numerical method}

In hadron-hadron collisions the individual 
contributions to an infrared-safe observable at next-to-leading order with $n$ final state particles 
can be written in a condensed notation as
\bq
\langle O \rangle^{\mathrm{NLO}} 
 = 
 \int\limits_{n+1}O_{n+1} \; d\sigma^{\mathrm{R}}
 +
 \int\limits_{n}O_{n} \; d\sigma^{\mathrm{V}}
 +
 \int\limits_{n}O_{n} \; d\sigma^{\mathrm{C}},
\eq
where $d\sigma^{\mathrm{R}}$ denotes the real emission contribution, 
which corresponds to the square of the tree amplitude with $(n+3)$ partons $|{\cal A}_{n+3}^{(0)}|^{2}$, 
$d\sigma^{\mathrm{V}}$ denotes the virtual contribution, 
which corresponds to the interference term of the renormalised one-loop amplitude 
with the tree amplitude $2{\mathrm{Re}}({\cal A}_{n+2}^{(0)^{*}} {\cal A}_{n+2}^{(1)})$, 
and $d\sigma^{\mathrm{C}}$ subtracts initial state collinear singularities. 
Each term is separately divergent and only their sum is finite. 
It is common practice to use the subtraction method in order to render the real emission contribution finite.
Within the numerical method we take the subtraction method one step further and use it as well for the virtual part.
With the help of suitable subtraction terms we can write the NLO contribution as
\bq
 \langle O \rangle^{\mathrm{NLO}} 
 & = &
 \langle O \rangle^{\mathrm{NLO}}_{\mathrm{real}}
 +
 \langle O \rangle^{\mathrm{NLO}}_{\mathrm{virtual}}
 +
 \langle O \rangle^{\mathrm{NLO}}_{\mathrm{insertion}}.
\eq
Each of the three terms on the right-hand side is individually finite and can therefore be computed with Monte Carlo techniques.
The three terms are given by
\bq
 \langle O \rangle^{\mathrm{NLO}}_{\mathrm{real}}
 & = &
 \int\limits_{n+1} \left(O_{n+1} \; d\sigma^{\mathrm{R}} - O_{n} \; d\sigma^{\mathrm{A}}\right),
 \nonumber \\
 \langle O \rangle^{\mathrm{NLO}}_{\mathrm{virtual}}
 & = &
 \int\limits_{n+\mathrm{loop}} \left(O_{n} \; d\sigma_{\mathrm{bare}}^{\mathrm{V}} - O_{n} \; d\sigma^{\mathrm{L}} \right),
 \nonumber \\
 \langle O \rangle^{\mathrm{NLO}}_{\mathrm{insertion}}
 & = &
 \int\limits_{n} 
  \Big( O_{n} \; d\sigma_{\mathrm{CT}}^{\mathrm{V}}
  + O_{n} \int\limits_{\mathrm{loop}} d\sigma^{\mathrm{L}}
  + O_{n} \int\limits_{1} d\sigma^{\mathrm{A}}
  + O_{n} \; d\sigma^{\mathrm{C}}
 \Big).
\eq
Here, $d\sigma^{\mathrm{A}}$ denotes the subtraction term for the real emission part, $d\sigma^{\mathrm{L}}$ denotes the subtraction term
for the virtual part. We also separated the renormalised virtual contribution $d\sigma^{\mathrm{V}}$ 
into a bare part and a counter term:
\bq
 d\sigma^{\mathrm{V}} & = & d\sigma_{\mathrm{bare}}^{\mathrm{V}} + d\sigma_{\mathrm{CT}}^{\mathrm{V}}.
\eq
There are several advantages of this approach:
First, for $\langle O \rangle^{\mathrm{NLO}}_{\mathrm{virtual}}$
we can combine the integration over the phase space of the final state particles with the integration over the loop momentum
into one Monte Carlo integration.
We recall that the Monte Carlo integration converges independently of the dimensionality of the integration region.
Secondly, the scaling behaviour of each integrand evaluation with respect to the number $n$ of external particles behaves as $n^3$, when working with cyclic-ordered primitive amplitudes.
We recall that the integrand of a one-loop amplitude is a tree-like object. This is most easily seen by cutting the loop open at one position.
In this context it is also important that the subtraction terms for the virtual parts can be computed efficiently.
This is indeed the case: For cyclic-ordered primitive one-loop amplitudes the infrared subtraction terms are very simple
and proportional to the corresponding tree amplitudes, whereas the ultraviolet subtraction terms are easily computed recursively from
propagator and vertex subtraction terms \cite{Assadsolimani:2009cz,Assadsolimani:2010ka,Becker:2010ng,Becker:2012aq}.

There is a second important ingredient within the numerical method:
The virtual infrared and ultraviolet subtraction terms ensure
that the integration over the loop-momentum gives a finite result and can therefore be performed in four dimensions.
However, this does not imply that the integration over the loop momentum can be performed in the real domain.
For real values of the loop momentum there is still the 
possibility that some of the loop-propagators go on-shell.
These singularities are avoided by a deformation of the integration contour into the complex plane.
Therefore we shift the integration contour into the complex space $\mathbb{C}^{4}$, 
where the integration contour must be chosen such that whenever possible the poles of the propagators are avoided. 
We set
\bq
 k & = &
 \tilde{k} + i \kappa\left(\tilde{k}\right),
\eq
where $\tilde{k}^{\mu}$ contains only real components. After the deformation our one-loop integral reads
\bq
I=
\int\frac{d^{4}k}{(2\pi)^{4}}
\frac{R(k)}{\prod\limits_{j=1}^{n}\left(k_{j}^{2}-m_{j}^{2}\right)}
=
\int\frac{d^{4}\tilde{k}}{(2\pi)^{4}}\left|\frac{\partial k^{\mu}}{\partial \tilde{k}^{\nu}}\right|
\frac{R(k(\tilde{k}))}{\prod\limits_{j=1}^{n}\left(\tilde{k}_{j}^{2}-m_{j}^{2}-\kappa^{2}+2 i \tilde{k}_{j}\cdot\kappa\right)},
\eq
where we integrate over the four real components in $\tilde{k}^{\mu}$ and $R(k)$ is a holomorphic function in the integration domain of interest. 
To match Feynman's $+i\eps$-prescription we have to construct the deformation vector $\kappa$ 
such that $\tilde{k}_{j}\cdot\kappa \geq 0$ whenever $\tilde{k}_{j}^{2}-m_{j}^{2}=0$,
and the equal sign applies only if the contour is pinched \cite{Gong:2008ww,Becker:2010ng,Becker:2012aq,Becker:2012nk,Becker:2012bi}. 
If the contour is pinched the singularity is either integrable by itself or there is a subtraction term for it. 
The loop integral is independent of the choice of the contour, as long as no poles are crossed.
However, the choice of the contour has a direct impact on the Monte Carlo integration error.

The Monte Carlo integration is in general more challenging for the contribution
$\langle O \rangle^{\mathrm{NLO}}_{\mathrm{virtual}}$ as compared to the contribution
$\langle O \rangle^{\mathrm{NLO}}_{\mathrm{real}}$.
The reason for this can be deduced from the underlying unsubtracted quantity.
In the real emission contribution we have the square of a tree amplitude, which is always positive.
In the virtual part we have the interference term of a one-loop amplitude with a tree amplitude, which in general
will be oscillating.
In simple terms we are facing an integral of the form
\bq
 I & = & \int\limits_0^1 dx \; \left[ c + A \sin\left(2\pi x \right) \right],
 \;\;\;\;\;\;
 A \gg c.
\eq
A naive Monte Carlo integration will lead for $A \gg c$ to large Monte Carlo integration errors.
A solution is given by the method of antithetic variates: One combines the evaluation of the integrand at $x$ and $(1-x)$.
Our strategy is therefore as follows: We first identify integration regions, where oscillations might be large.
Not surprisingly, these regions are the ultraviolet and infrared regions.
We then split the loop integration into several channels (one ultraviolet channel and $n$ infrared channels, each infrared channel
corresponds to one of the $n$ loop propagators). Within each channel we can choose an appropriate coordinate system, 
where the choice of ``good'' variables for the method of antithetic variates is simple \cite{Becker:2012aq}.
For the ultraviolet channel, we can even choose a different contour.
We emphasize that this procedure respects the universality of our approach, it
does not depend on the specific process under consideration.

\section{General improvements}

In addition to improvements closely related the numerical method, we also developed efficiency improvements, which can
be used not only within the numerical method, but are also suitable to other approaches like the unitarity method
or an approach based on Feynman graphs.

We first mention the extension of the dipole formalism towards random polarisations.
The matrix elements are usually calculated from helicity amplitudes and involve a sum over $2^n$ helicity configurations
for $n$ external particles, under the assumption that each external particle has two helicity states.
The exponential growth of $2^n$ spoils the desirable polynomial behaviour.
The cost factor $2^n$ can be eliminated as follows:
One first introduces random polarisations \cite{Draggiotis:1998gr}
\bq
 \eps_\mu(\phi) & = & e^{i\phi} \eps_\mu^+ \; + \; e^{-i\phi} \eps_\mu^-,
\eq
where $\eps_\mu^\pm$ are the polarisation vectors of the helicity eigenstates.
In a second step one replaces the summation over the helicity states by an integration over the angle $\phi$:
\bq
 \sum\limits_{\lambda=\pm} {\eps_\mu^\lambda}^\ast \eps_\nu^\lambda 
 & = & 
 \frac{1}{2\pi} \int\limits_0^{2\pi} d\phi \; {\eps_\mu(\phi)}^\ast \eps_\nu(\phi).
\eq
This works straightforwardly for the Born and the virtual part.
However, for the real emission part the subtraction terms are usually spin-summed and thus non-local in $\phi$.
In order to use the method of random polarisations also for the real emission part, 
we extended the dipole formalism \cite{Catani:1997vz,Phaf:2001gc,Catani:2002hc}
to ensure that all singularities are subtracted locally not only in phase space, 
but also with respect to the helicity angles $\phi$ \cite{Gotz:2012zz}.

A second improvement concerns the colour decomposition at one-loop.
It is convenient to 
organise the computation of the one-loop amplitude as a sum over smaller pieces, called primitive amplitudes.
Primitive amplitudes are gauge invariant, have a fixed cyclic ordering of the external legs
and a fixed routing of the fermions through the loop.
For amplitudes with more than one quark-antiquark pair the decomposition of the full amplitude into primitive
amplitudes is non-trivial.
However, for multi-parton final states amplitudes with several quark-antiquark pairs are an essential ingredient.
One possible algorithm is based on Feynman diagrams and the solution of a system of linear equations \cite{Ellis:2008qc,Ellis:2011cr,Ita:2011ar,Badger:2012pg}, but a method which avoids Feynman diagrams and the need to solve a large system
of linear equations is clearly preferred.
In \cite{Reuschle:2013qna} we showed that the decomposition into primitive amplitudes can be obtained directly through shuffle relations.
This method is also discussed in these proceedings \cite{Reuschle:2014qwa}.

\section{First results for $pp \rightarrow Z  + 5 \; \mbox{jets}$}

We now present first preliminary results for the NLO QCD corrections to the process $pp \rightarrow Z  + 5 \; \mbox{jets}$.
For our analysis we use the following set of cuts, which was also used by the Blackhat collaboration \cite{Ita:2011wn}:
We include the decay $Z, \gamma^\ast \rightarrow e^+ e^-$ and require for each lepton $l$
\bq
 p^\perp_l > 20 \; \mathrm{GeV},
 & &
 \left| \eta_l \right| < 2.5.
\eq
For the invariant mass of the lepton pair we require
\bq
 66\;\mathrm{GeV} < m_{l\bar{l}} < 116\;\mathrm{GeV}.
\eq
The jets are defined by the anti-kt-algorithm \cite{Cacciari:2008gp} with $R=0.5$.
For $Z,\gamma^\ast + n \; \mathrm{jets}$ we consider inclusive jet production, i.e. we require at least $n$ jets with
\bq
 p^\perp_{\mathrm{jet}} > 25 \; \mathrm{GeV},
 & &
 \left| \eta_{\mathrm{jet}} \right| < 3.
\eq
We consider collisions at $\sqrt{s}=7\;\mathrm{TeV}$. 
We use the CTEQ6M pdf set at NLO and the CTEQ6L1 pdf set at LO \cite{Stump:2003yu}.
The renormalisation and factorisation scale are chosen on a per-event basis. The nominal choice is
\bq
 \mu_{\mathrm{R}} = \mu_{\mathrm{F}} =
 \frac{1}{2} {H^\perp}'
\eq
with
\bq
 {H^\perp}' & = & E_Z^\perp + \sum\limits_j p_j^\perp,
 \;\;\;\;\;\;
 E_Z^\perp = \sqrt{ m_Z^2 + \left( p^\perp_{l\bar{l}} \right)^2}.
\eq
The sum runs over all final state partons.
For $\alpha_s$ we take at LO the QCD parameter $\Lambda_{\mathrm{LO}}^{(5)}=165.2 \;\mathrm{MeV}$, corresponding to $\alpha_s^{\mathrm{LO}}(m_Z)=0.130$, while
at NLO we take $\Lambda_{\mathrm{NLO}}^{(5)}=226.2 \;\mathrm{MeV}$, corresponding to $\alpha_s^{\mathrm{NLO}}(m_Z)=0.118$.
These values are consistent with the corresponding pdf sets.

The preliminary results are in the leading-colour approximation.
The squared matrix elements can be expanded in $\alpha_s$ and in $1/N_c$, where $N_c$ denotes the number of colours.
For $pp \rightarrow e^+ e^- + n \;\mbox{jets}$ we have for the leading terms
\bq
 \mathrm{LO}, \mathrm{lc} & \sim & N_c \left( \alpha_s N_c \right)^n,
 \nonumber \\
 \mathrm{NLO}, \mathrm{lc} & \sim & N_c \left( \alpha_s N_c \right)^{n+1}.
\eq
The leading-colour contribution is entirely given by the amplitude $0 \rightarrow q \bar{q} e^+ e^- + n g$.
We remark that for $\alpha_s$ and for the evolution of the pdf's we take the full QCD running,
including sub-leading colour contributions.
The mismatch is formally beyond the leading-colour approximation and will disappear once the full-colour computation is available.
In the leading-colour approximation we obtain the preliminary results
\bq
 \sigma_{\mathrm{LO,lc}} & = & 0.138 \pm 0.009 \; \mathrm{pb},
 \nonumber \\
 \sigma_{\mathrm{NLO,lc}} & = & 0.161 \pm 0.113 \; \mathrm{pb}.
\eq
The large Monte Carlo integration error of the NLO result will go down with higher statistics.
The numbers above correspond to a Monte Carlo run of three days on a cluster with 200 cores.


\bibliography{/home/stefanw/notes/biblio}
\bibliographystyle{/home/stefanw/latex-style/h-physrev5}

\end{document}